\documentclass{PoS}
\usepackage{epsfig}

\title{Excited charmonium states from Bethe-Salpeter equation}

\ShortTitle{Excited charmonium}

\author{\speaker{Vladimir Sauli}\\
        Department of Theoretical Physics, NPI, Rez near Prague\\
        E-mail: \email{sauli@ujf.cas.cz}}

\author{Pedro Bicudo\\
        Institute Superior Tecnico, FCTP, Lisboa\\
        E-mail: \email{pedro.bicudo@gmail.com}}

\abstract{We solve the Bethe-Salpeter equation for a system of a heavy quark-antiquark pair  interacting with a screened linear confining potential. 
First we show the spinless QFT model is inadequate and fail to describe even gross feature of the quarkonia spectrum.
In order to get reliable description the spine degrees of freedom has to be considered. Within the approximation employed we reasonably reproduce  known radial excitation of vector charmonium. The BSE favors relatively large string breaking scale $\mu\simeq 350MeV$ .  
Using free charm quark propagators we observe that $J/\Psi$ is the only charmonium left bellow naive quark-antiquark threshold $2m_c$, while the all excited states are situated above this threshold. Within the  numerical method we overcome obstacles related with threshold singularity  and discuss the consequences of the  use of  free propagators for calculation of excited states above the threshold.}

\FullConference{International Workshop on QCD Green's Functions, Confinement and Phenomenology,\\
		September 05-09, 2011\\
		Trento Italy}

\newcommand{\be}{\begin{equation}}
\newcommand{\ee}{\end{equation}}
\newcommand{\bea}{\begin{eqnarray}}
\newcommand{\eea}{\end{eqnarray}}

\newcommand{\nn}{\nonumber}
\newcommand{\ep}{i\epsilon}

\begin{document}

\section{Introduction}

Excited meson spectroscopy is a keystone experimental output, which
is essential for understanding of quark-antiquark interaction. 
The dynamics of its constituents is dominantly  driven by solely known strongly interacting quantum field theory-
quantum chromodynamics.  The explanation of  confinement of constituents and all other colored objects is one of the great challenge of theory of strong interaction. The  confining forces between quark antiquark inside mesons should lead           
a large degeneracy which emerges from the spectra of the angularly and radially excited resonances.
As reported in \cite{Bugg:2004xu,Afonin:2006vi,Glozman:2007ek,fig1afon} such degeneracy is observed in $p \bar p$ annihilation by the Crystal Ball collaboration at LEAR 
in BERN \cite{Aker:1992ny}. Similar can be deduced  from heavy quarkonia production in $e^{+}e^{-}$ anihilations.
Recently, BABAR, Belle and BESS experiments continue collection of various meson experimental data. 

Heavy quarkonia, as relatively simple bound state systems made of the quark and antiquark of the same flavour, attract many theoretical and experimental attentions \cite{QUARKONIA}. 
 Following ideas of Ref. \cite{Wilson:1974}, confinement in heavy flavor hadron sector has typically been associated with a linearly rising potential between constituents \cite{Eichten:1978}. The spin degeneracy observed in heavy quarkonia spectrum  tell us that 
the main confining part of the interaction should be largely spin independent. A various lattice fits leaded to various predictions for the potential between quark-antiquark states. A well known is  Cornell parameterization of the static Wilson loop derived potential \cite{Bali2001,Greensite2003},
\be 
V(r)=-\alpha/r+\sigma r
\ee
which appeared to be suited for description of the first few excited mesons.

In the absence of dynamical quarks the nonrelativistic static potential shows up a linear asymptotic  of the interquark potential.
Recently in Refs. \cite{Licha2009,Chao1,Vento2011} it has been found that the meson spectroscopy is better described  by "confining"  potential which is bounded from above.  The potential is flattened in large distance due to the string breaking and the screening mass effectively describes the inverse of string breaking range. To this point the exponential potential can be regarded as a screened version of the linear potential in the region important for a  few lowest states. The screening effect - the flatness of linearly rising potential  arises form the string breaking scenario: including light quarks into the game then the creation of the quark-antiquark pairs, i.e. pions and other light mesons is energetically favorable. The screening is necessarily included to explain observed high radially excited heavy mesons- without any doubt, the spectrum deviates from the linear Regge trajectory. Recall, the deviations from the linear Regge trajectory is expected in the light meson sector as well \cite{PETR}.

On the other side screening of Coulomb potential is more subtle matter, which could be related with soft gluon mass generation \cite{PBC1,PBC2} through the Yang-Mills Schwinger mechanism. Actually the dynamical gluon mass  generation is  suggested by finiteness of the lattice  gluon propagator in the deep infrared, which fact is  apparently pronounced in the recent lattice data \cite{Mendes}. Both  scales, the first given by the inverse of soft gluon mass and the second- the screening mass characterizing string breaking, are independent in principle, albeit one naturally expects their similar size $\simeq \Lambda_{QCD}^{-1}$.

Using two models based on Bethe-Salpeter equation we study the effect of retardation in the slopes of radially excited vector mesons. In one of the  model we completely ignore the spin degrees of freedom in order to see the effects of retardation and effect of off-shellness of the constituents scalar propagators  in fully Lorentz invariant BSE approach.  The second model is based on  quite standard assumptions- the (anti)quarks have spin $1/2$ with the scalar confining part of the interaction and effective vectorial one gluon exchange. 

For bound states which lie above the naive quark-antiquark threshold, the BSE kernel becomes singular, which causes that  many usual numerical treatments fails and/or become impossible in practice.
To authors knowledge, theoretical explanation of plethora  experimentally identified heavy quarkonia masses has been obtained only in relativistic quantum mechanic \cite{RQM} or by solving  various phenomenological 3D reduction of BSE \cite{TJON1993,INSTANT}. The main motivation is to fill this missing gap and thus present our BSE study of above threshold excited bound states as a necessary step towards our better understanding of excited mesons in the nature.

\section{Model with spinless quarks}

In traditional quantum mechanical wisdom the spin interaction between quark and antiquark inside the mesons is necessary for relatively small  hyperfine splitting and thus neglection of  the quarks spin one should reproduce Regge trajectory for orbitally and radially excited mesons as well. Such picture has been confirmed in refs \cite{DUKASI93,KANESI2001} where  spinless quarks has been considered.

In the first model we just consider spinless quarks interaction through the relativistic invariant generalization of exponential and Coulomb(Yukawa) potential (see the Section next for the discussion of the nonrelativistic limit). To accomplish this we solve the following BSE  
  
\be
\Gamma(p,P)=-i\int\frac{d^4k}{(2\pi)^4}  V(k,p,P)G^{[2]}(k,P)\Gamma(k,P) 
\ee
where  $G$ is  two body propagators
\bea
G^{[2]}(k,P)&=&D(k+P/2,m_c^2)D(-k+P/2,m_c^2) \nn
 \\
D(k,M^2)&=&\frac{1}{k^2-m_c^2-i\ep} \nn
\eea
where $m_c$ is constituent quark mass
and the interaction  kernel is chosen such that

\bea 
V&=&V_s+V_v \nn
 \\
V_s(q)&=&\frac{C}{(q^2-\mu^2)^2} \nn
\eea
and effective OGE
\be \label{kernel2}
V_v(q)= \frac{g^2}{q^2-\mu^2_g} \nn
\ee
with $q=k-p$ and where $\mu=\mu_g\simeq (0.1-0.3)m_c$ for simplicity  ($m_c$ is the constituent charm quark mass). 
Recall here that BSE considered here (and in fact in the next section as well) thus represent  generalization of nonrelativistic  model considered 
\cite{Chao1,Licha2009} where the potential (\ref{mechanic}), complemented by coulombic term and constant positive shift $\sigma/\mu$, provides quantum mechanical model describing experimentally known charmonium spectrum.

The BSE has been solved after performing Wick rotation and integration over the 3d space angles. It transforms the 4-dimensional BSE into the 2-dimensional one which has been solved by the method of the iterations
by the similar method described in the recent paper \cite{SAULI2011}.
The  2d BSE we have actually solved reads

\bea \label{BSEscalar}
\Gamma(p,P)&=& \int_{-\infty}^{\infty}{d k_4} \int_{0}^{\infty}d {\bf k}\, \Gamma(k,P) G_2\left[K_s+K_v\right] \, ;
 \\
K_s&=&\frac{C}{4\pi^3}\frac{\bf{k^2}}{\left[k^2_E+p^2_E+2k_4p_4+\mu^2\right]^2-4\bf{k^2}\bf{p^2}}
\nn
\\
K_v&=&\frac{g^2}{4\pi^3}\frac{\bf k}{\bf p}
\ln\left[\frac{k^2+p^2-2k_4p_4-2|{\bf k}||{\bf p}|+\mu^2}{k^2+p^2-2K_4p_4+2|{\bf k}||{\bf p}+\mu^2}\right]
\nn \, \,
G_2^{-1}=(k_E^2-P^2/4+m_c^2)^2+k_4^2P^2 \nn \\
\eea

Let us only mention here that the kernel of  BSE becomes singular for the bound states heavier then sum of constituents, which requires certain numerical  care when is solved numerically.

For expected values of the parameters $C,g$ the ground state appears near , but allays bellow the threshold $3GeV$ . This  s-state corresponds with degenerate experimental partners of $\eta_c$ and $J\psi$ mesons. We attempted to fit the distribution of other known s-states states, e.g. by adjusting $C,g,\mu$ to get $\psi(2S)$,$\psi(3S)$, etc... 
In order to ensure the numerical stability we also varied integration volume (in p-space) and the number of points as well. In order to get optimal density of integration points the upper boundary $\Lambda$ is introduced in the momentum integrations. Let us stress, the wave function  is exponentially decreasing and we always take  $\Lambda$ such that cut contribution is largely negligible ensuring that the main role of $\Lambda$ is to adjust the density of integrations points and not to regulate  the finite integrals. More precisely, for  the variable ${\bf k}$ we map single integration interval $(0,1)$ into  $(0,\Lambda)$ by simple rescaling (and   $(-1,1)$ into  $(-\Lambda,\Lambda)$   for variable $k_4$).

\begin{figure}[t]
\begin{center}
\centerline{  \mbox{\psfig{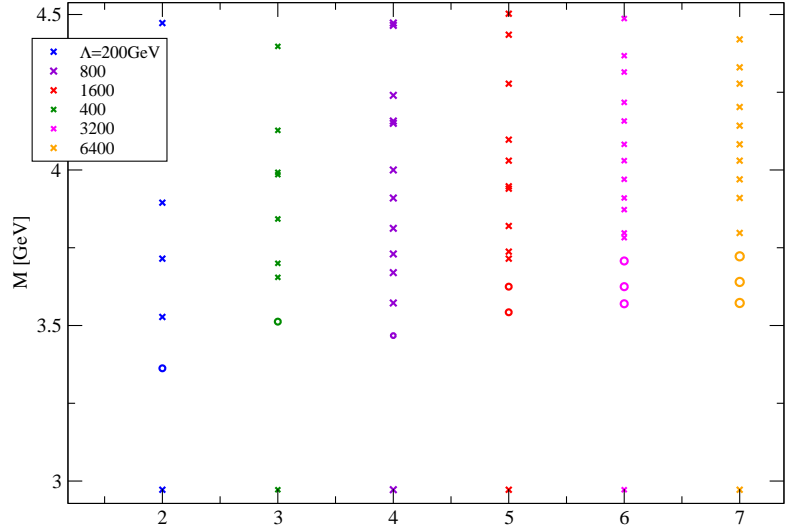}} }
\label{regge1}
\caption{Solution of BSE for $C =6 GeV^{4}$, $g^2=0.5*(2\pi)^2GeV^2,
\mu=0.15GeV, m_c=1.5GeV $ and various cutoff $\Lambda$. Volume increases from the left to the right}
\end{center}
\end{figure}

\begin{figure}[t]
\begin{center}
\centerline{  \mbox{\psfig{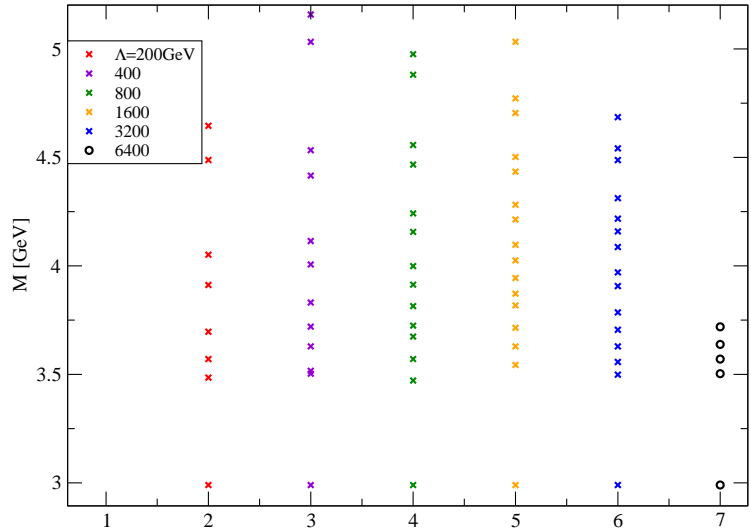}} }
\label{regge2}
\caption{Showing instability. The same as in fig.1 but for $C =1.6 GeV^{4}$ . }
\end{center}
\end{figure}

The results are shown in figures \ref{regge1} for different volumes and its general features do not change considerably when changing the parameters
in a range  $\pm 30\%$.
While it is not difficult to fit two or three states we have failed to reasonably describe the   experimentally  known $s-states$ as a whole, the obtained discrepancy is tedious since  one allays gets the distribution of the states considerably dense then one can deduce  from the known  experiments. For instance, decreasing $C$ does not lead large decrement of the density of the states , instead of, the numerical stability is lost, see Fig. \ref{regge2}. Increasing $g^2$ one can shift the intercept, but the rest of the energy levels are only slightly changed (also the numerical value is limited from the above by some critical value, above which the BSE spectrum become continuous).

\section{Model II, Vector BSE for mesons}

In principle the framework of Schwinger-Dyson and Bethe-Salpeter equations in QCD should offer  unique  fully Poincare invariant  generalization of well known quantum mechanical picture of quarkonia, however a solution is always incomplete due to the truncation of the equations system. 
For instance how to truncate the set of equations in  order to reproduce Wilsonian loop  potential still remains unclear.   Due to this fact we rather phenomenologically estimate what should be the form of the  BSE kernels here. 

In what follows we consider  "confining" interaction kernel of the form
\be  \label{kernel}
V_s(q)=\frac{C}{(q^2-\mu^2)^2} \, ;
\ee
which couples as a scalar in the quark-antiquark scattering kernel in the meson BSE.
The second considered part of the interaction kernel has the Dirac decomposition
identical with the one gluon exchange $\simeq \gamma^{\mu}_{\alpha\beta}V_v\gamma^{\nu}_{\alpha^{,}\beta^{,}}$.  

For clarity we write down the BSE completely here

\bea \label{BSEcelkove}
S^{-1}(q+P/2)\chi(p,q)S^{-1}(q+P/2)&=&-i\int \frac{d^4k}{(2\pi)^4}\gamma_{\mu}\chi(k,P)\gamma_{\nu}G^{\mu\nu}(k-q)
\nn \\
&-&i\int \frac{d^4k}{(2\pi)^4}\chi(k,P)V_s(k-q) \, ;
\nn \\
G^{\mu\nu}(k-q)&=&g^{\mu\nu}V_v=\frac{g^2g^{\mu\nu}}{(k-q)^2-\mu_g^2}  \, ; 
\eea

In words, scalar functions $V_s$ and $V_v$ are Poincare invariant generalization of quantum mechanical potentials.
Here, clearly $G^{\mu\nu}$ represents Gluon propagator in Feynman like gauge, where the effective soft gluon mass $\mu_g$ has been introduced.  
Like in the previous scalar case, the double pole scalar interaction $V_s$ leads to regular exponential potential in the position space.  
Actually,  in heavy quark limit one can consider three dimensional potential

\bea\label{pot2}
V_s^{QM}(\vec{k}) 
&=& \int_0^\infty  dr  \, { 4 \pi r \sin( k r ) \over k }  \, V( r)  
\nonumber \\
&=&  \sigma { - 8 \pi \over ( {\bf \vec{k}^2} + \mu^2)^2   } \ , 
\eea
where the potential in position space reads 
\be
V(r)=- \sigma {e^{- \mu \, r} \over \mu} \, .
\label{mechanic}
\ee

 $S$ in Eq. (\ref{BSEcelkove}) stands for charm quark propagator, which in the simplest approximations reads
\be
S^{-1}(l)=\not l-m_c\, , 
\ee  
and $\chi_V$ represents  the Bethe-Salpeter wave function which  has the general form:
\bea
\chi_V(q,P)=\not\epsilon \chi_{V0}+ \not P \epsilon.q\chi_{V1}+ \not q \epsilon.q\chi_{V2}+\epsilon.q\chi_{V3}
\nn \\
+ [\not\epsilon,\not P]\chi_{V4}+[\not\epsilon,\not q]\chi_{V5}+[\not q,\not P]\chi_{V6}+i\gamma_5\not t\chi_{V7} \, \, ,
\eea
with $t_{\mu}=\epsilon_{\mu\nu\alpha\beta}q^{\nu}P^{\alpha}\epsilon^{\beta}$,
$\epsilon^2=-1$, $\epsilon.P=0$.

Munczek and Jain \cite{JAMU} shown that $V0$ component is   dominant for the all ground state  mesons, which  dominance is particular for mesons made form  heavy flavor (anti)quarks. We assume the same apply for excited states as well  and we neglect all other components in  presented study.  
 
Within the approximation the Bethe-Salpeter equation in the rest frame reads
\be \label{BSE1}
\chi_{V0}(p_E,P)=\frac{-p^2_E-m^2-P^2/4}{(-p^2_E-m^2+M^2/4)^2+q^2_4M^2} I_0 \,\, ,
\ee
where
\bea
I_0&=&-\int\frac{d^4k_E}{(2\pi)^4}\left[-2V_v+V_s\right] \chi_{V0}(k_E,P)  \, \, ;
\nn \\
V_v&=&\frac{g^2}{q^2_E+\mu^2} \, ;
\nn \\
V_s&=&\frac{C}{(q^2_E+\mu^2)^2} \, ,
\nn \\
\eea
where we have performed Wick rotation into the Euclidean space, for Euclidean momenta $k_E=(k_4,\vec{k})$, $k_E^2=k_4^2+{\bf k}^2$, while the total momenta is kept timelike $P^2=-P^2_E=M^2$ as required for the bound states.
Again an extra  problem arises for bound state which are heavier then sum of constituents quark masses. 
As we are using propagators with single real poles, the BS wave function becomes singular since being proportional to  the product of quark propagators.
For the vertex function, the  threshold like singularity should  appear for the solution with $P^2>4m_q^2$.
Due to the presence of the kernel singularity more or less standard matrix methods \cite{BLAKRA2010} fail  since the inversion of numerical matrices is not possible.  Also we do not explore more or less conventional  expansion into the orthogonal polynomials which loses its efficiency  when, as one expects, relatively large number of polynoms is necessary. Instead of, we rather solve the full two dimensional integral equation by the method of simple iterations.
For this purpose we discretize $P^2$ and step by step we are looking for the solution of the BSE with given $P_i^2$. Performing several hundred iterations for each $P_i^2$ we identify the solutions as those for which difference between iterations vanishes. 

Without loosing the generality the BSE can be transformed into the 2dim integral homogeneous equations. It requires normalization,  Instead of  using physical one, 
to achieve a good numerical stability of the iteration process we implement normalization condition through an auxiliary function $\lambda(P)$.
We choose  $\lambda(P)$ such that it   makes BSE nonlinear but mainly numerically stable and such that  the BSE solution has been identified when $\lambda(P)=1$. Simultaneously the integrated difference $\sigma$ between  consecutive iterations must vanish at the same time. We found that these two conditions happen simultaneously, while for other values of  parameters $P,\lambda(P)\neq 1$ the numerics do not provide vanishing difference between iterations.  For interested readers the details of the numerics is described in the paper \cite{SAULI2011}.

\begin{figure}[t]
\begin{center}
\centerline{  \mbox{\psfig{figure=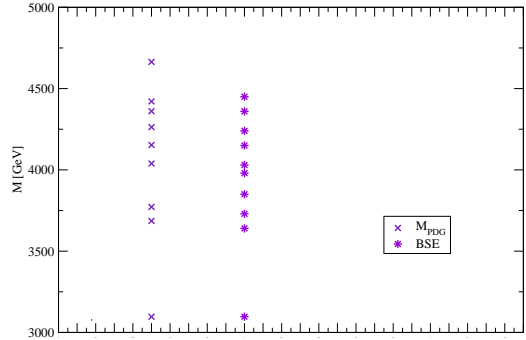,height=8.0truecm,angle=270}} }
\label{huf}
\caption{Comparison of BSE solution with PDG data for vector charmonium.}
\end{center}%
\end{figure}

\begin{table}[b]
\begin{center}
\small{
\begin{tabular}{|c|c|c|c|} \hline \hline $M_{64}$ & $M_{\epsilon,96} $ &
$PDG$  & $n,l $   \\
\hline \hline
3097 & 3097 & 3097 & 1s \\
\hline
-& 3639   & 3686 & 2s \\
\hline
3724 & 3733 & 3772 & 1d\\
\hline
3866 & 3855 & -- &  \\
\hline
3998&  3980 & -- &  \\
\hline
4053 & 4033 & 4039 & 3s \\
\hline 
4160& 4149 & 4153 & 2d\\
\hline 
-& 4243 & 4263 & 4s\\
\hline
-& 4365 & 4361 & 3d \\
\hline
-& 4503 & 4421 & 5s \\
\hline \hline
\end{tabular}}
\caption[99]{Comparison  of calculated BSE spectrum with PDG data (third column). The first column represents preliminary results presented at the talk at QCD-TNT2011. They have been calculated on the grid with $\Lambda=1000 GeV$ and 64*128 number  of points. The second column represents the new results published in \cite{SAULI2011} which has been obtained with 96*192 points. Quantum numbers correspond with  assumed quantum mechanical assignment \cite{Vento2011}.
\label{buchta}}
\end{center}
\end{table}

  As in the previous case of spinless BSE the  numerical convergence of $\lambda\rightarrow 1$ and the resulting numerical solution itself crucially depend on the density of integration point. Therefore, similarly to the previous treatment we introduce cutoff $\Lambda$ for the purpose.

For clarity  we restrict ourself to the dynamics of charmonia in this paper. The numerical values of the models are the following
\bea 
C&=&5.418 GeV^2\, \, ; \alpha_s=g^2/(4\pi)=0.2 \\ 
\nn m_c&=& 1.5615 GeV ; \mu=\mu_g=364.35 MeV  \, 
\eea
 and we use the mass of $J/\Psi$ meson  to fine tune the correct scale at the end  ($m_c=1.5 \, ;\,  C=5 $ in units where $M(J/\Psi )=2975$).     
 The results presented in the  Tab. \ref{buchta} have been calculated for 
$\Lambda=1000GeV$ where the numerical stability has been (almost) achieved.
We also compare with the experimental data  in the Tab. \ref{buchta} and Fig. \ref{huf}. 

  As one can see in the Table \ref{buchta}. there are two more BSE solutions with masses situated in between $\psi^{''}$ and $\psi^{'''}$, while  the rest of calculated excited states quite nicely agree with the data. We have not found possible way (by varying parameters $C,\alpha,mu,m_c$) to exclude these two additional states from the solutions.

\section{Discussion of the results}

We have formulated two BSEs models for vector quarkonia.
Ignoring the spin degrees of freedom and treating quarks as spinless particles is not justified approximation when using BSE.
While the presented numerical results do not represent ultimate numerical search we argue here that the retardation effects can be largely different in spinless and spin-incorporated BSE.
In the second model -with two aforementioned exceptions of two additional states- the resulting spectrum is comparable with the experiments whenever the experimental data are available.  The agreement between our results for higher states and the one measured in the experiments is impressive, remaining   difference between theory and experiments is due to the approximations, e.g.  due to the interplay of quark and ($D,D*$) mesonic degrees of freedom, such  couple channel effects  are difficult to incorporate into the Bethe-Salpeter analysis presented here.

At given stage the question of confinement is beyond scope of the presented work, however we expect some  changes when confinement is correctly incorporated.
First of all, we expect the  quark propagators should not have a free particle pole and therefore the BSE could not posses ordinary threshold singularity in this case.  
In the paper we are dealing with BSE where the propagators  describe free -instead of confined- quarks. Therefore the threshold singularity unavoidably appears as an artefact here. The lowest lying excited charmonia are the one  closest to the naive quark-antiquark threshold and we naturally must expect
some defects in the calculated spectrum.  We argue two more states which appear are  the artefact of inappropriate usage of the free quark propagators. Numerical solution and an implementation of the confinement at this level  remains for future study.

Furthermore, we expect that the knowledge of off-shell behavior ($q_4$ dependence) of the BSE amplitudes can be important in various hadronic processes.
Due to this it is worthwhile to study not only the static property like the  spectrum here, but the cross sections  of a processes including studied mesons.


\end{document}